\documentclass[preprint,showpacs,showkeys,aps,prb]{revtex4}
\usepackage[dvips]{graphicx}% Include figure files

\begin{document}
\title{Shockwaves and Local Hydrodynamics;
Failure of the Navier-Stokes Equations}
\author{Wm. G. Hoover and Carol G. Hoover      \\
Ruby Valley Research Institute                  \\
Highway Contract 60, Box 598                    \\
Ruby Valley, Nevada 89833                       \\
}

\date{\today}

\pacs{47.40.-x, 05.20.Jj, 47.40.Nm, 47.10.ad}

\keywords{Shockwaves, Molecular Dynamics, Navier-Stokes, Compressible Flows}

\begin{abstract}
Shockwaves provide a useful and rewarding route to the nonequilibrium properties of
simple fluids far from equilibrium.  For simplicity, we study a strong
shockwave in a dense two-dimensional fluid.  Here, our study of 
nonlinear transport properties makes plain the connection between the
observed local hydrodynamic variables (like the various gradients and
fluxes) and the chosen recipes for defining (or ``measuring'') those
variables.  The {\em range} over which nonlocal hydrodynamic
averages are computed turns out to be much more significant than are
the other details of the averaging algorithms.  The results show
clearly the incompatibility of microscopic time-reversible cause-and-effect
dynamics with macroscopic instantaneously-irreversible models like the
Navier-Stokes equations.

\end{abstract}

\maketitle

\section{Introduction}
Leopoldo Garc\'ia-Col\'in has studied nonequilibrium fluids throughout
his research career. In celebrating his Eightieth Birthday we conform here
to his chosen field of study. Though Leo's approach is typically quite
general, looking for improvements on linear transport theory, he has studied
particular problems too.  A specially interesting and thought-provoking
study, with Mel Green, of the nonuniqueness of bulk viscosity\cite{b1},
emphasised the general problem of finding appropriate definitions for 
state variables far from equilibrium.  The magnitude of the bulk viscosity
gives the additional viscous pressure due to the compression
{\em rate}.  The pressure difference evidently depends upon the underlying
definition of the equilibrium reference pressure.  The reference pressure
itself in turn depends upon the choice between temperature and energy in
{\em defining} the reference state.  In the end, the same physics
results, as it always must; the valuable lesson is that many different
languages can be used to describe the underlying physics.  There is the
tantalizing possibility that some one approach is better than others.

In fact, temperature itself can have {\em many} definitions away from 
equilibrium\cite{b2}.  Away from equilibrium the {\em thermodynamic}
temperature would depend upon defining a nonequilibrium entropy -- and
there is good evidence that there is no such entropy. This is because
nonequilibrium distribution functions are typically fractal, rather
than smooth\cite{b3}.   The {\em kinetic} temperature, a measure of the
velocity fluctuation, becomes a {\em tensor} away from
equilibrium\cite{b2,b4}.
At low density this temperature is the same as the pressure tensor,
$P = \rho T$.  For
dense fluids the potential energy introduces {\em nonlocality}, complicating
the definition of constitutive averages.
The simplest of the many {\em configurational} temperatures\cite{b5,b6,b7}
depends on force
fluctuations, and so likewise has tensor properties.  Because
configurational temperature can be negative\cite{b8}
and because thermodynamic temperature is undefined away from
equilibrium, we focus our attention on kinetic temperature here.

%Figure 1 goes here
\begin{figure}
\includegraphics[height=8cm,width=6cm,angle=-90]{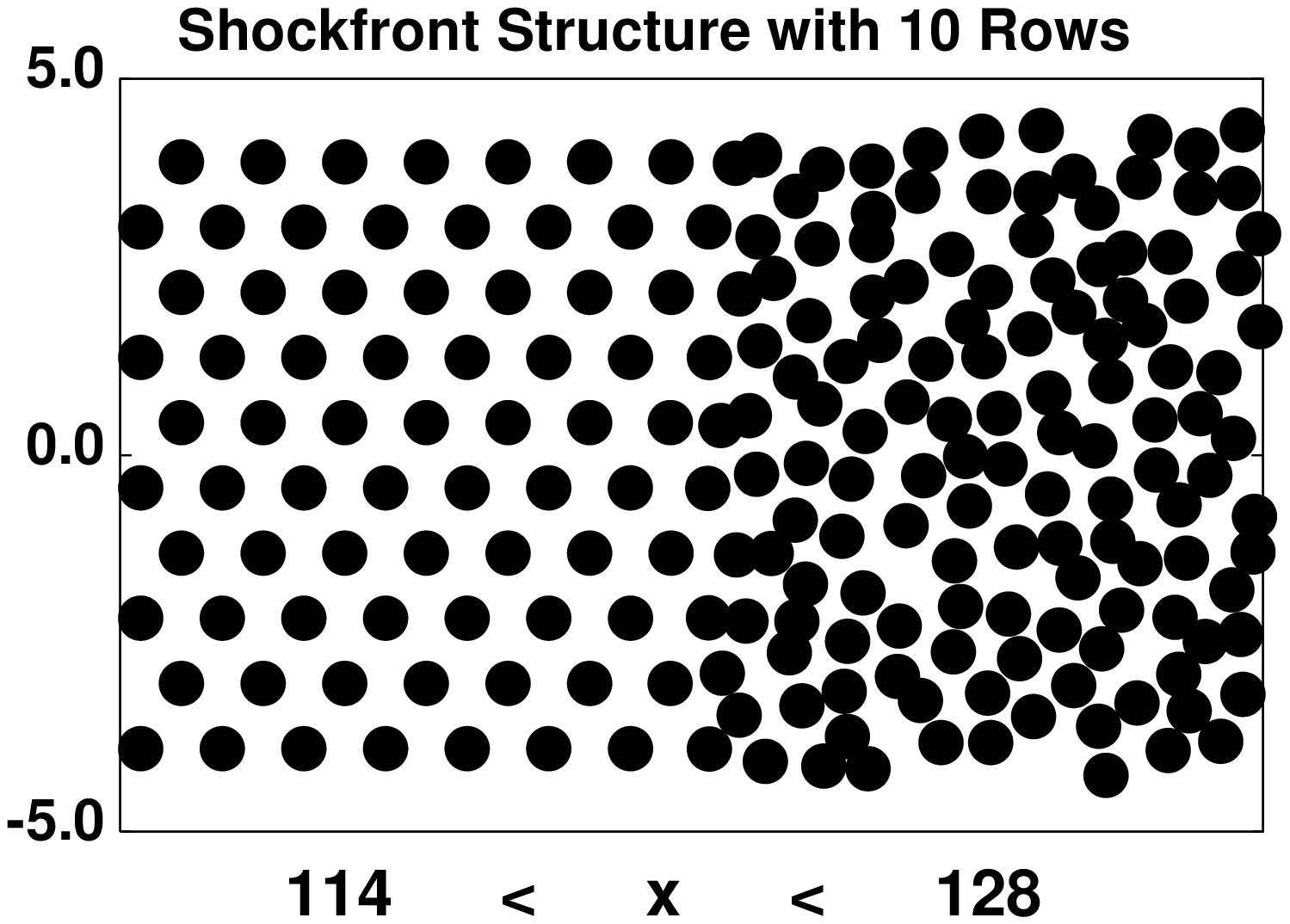}
\caption{
Closeup of a strong shockwave.  The cold stress-free solid on the left moves
to the right at twice the speed of the hot fluid, which exits to the right.
The boundaries in the vertical $y$ direction are periodic.
The pair potential is $\phi(r<1) = (10/\pi)(1-r)^3$.  The overall density change
is $\sqrt{4/3} \rightarrow 2\sqrt{4/3}$ and $u_s = 2u_p = 1.93$.  The system
height is $10\sqrt{3/4} \simeq 8.66$.
}
\end{figure}

Shockwaves are irreversible transition regions linking a ``cold'' and
a ``hot'' state\cite{b8,b9,b10,b11}.  Such a shock region contains
nonequilibrium gradients in density, velocity, and energy.  The
irreversible change from cold to hot takes place in just a few free paths
in a time of just a few collision times\cite{b11}.  The localized nature
of shockwaves makes them ideal for
computer simulation.  Their gross one-dimensional nature, illustrated in
Figure 1, makes it possible to compute {\em local} averages in a region
of width $h$.  Because $h$ is necessarily small it is evident that the
average values depend on it.  Thus the average temperature depends upon
both the underlying definition of temperature and additionally on
the details of the local averaging.

In this work we begin by describing molecular dynamics simulations,
for a strong, nominally stationary and one-dimensional shockwave,
in a two-dimensional fluid.  Next, we discuss
the Navier-Stokes description of such a wave and then set out to compare
the two approaches, focusing on the definition of local hydrodynamic
variables.  A close look at the momentum and heat fluxes shows clear
evidence for the incompatibility of the microscopic and macroscopic
constitutive relations. 

\section{The Microscopic Model System and a Continuum Analog}

We consider structureless particles of unit mass in two space dimensions
interacting with the short-ranged purely-repulsive pair potential,
$$
\phi(r<1) = (10/\pi)(1-r)^3 \ .
$$
As shown in Figure 1, particles enter into the system from the left,
moving at the shock velocity $u_s$.  Likewise, particles exit at
the right with a lower mean speed, $u_s - u_p = u_s/2$, where $u_p$ is the
``Particle'' or ``piston'' velocity.  The velocity ratio of two which
we choose throughout is consistent with
twofold compression.  We carried out series of simulations, all with a
length of 250 and the shock near the system center, with system widths
of from 10 to 160 rows.  Figure 1 shows a closeup of the center of such
a 10-row flow for the narrowest system width, $10\sqrt{3/4} \simeq 8.66$. 

To analyze the results from molecular dynamics one and two-dimensional
average values of the density, energy, pressure,
heat flux and the like were computed using the one- and two-dimensional
forms of Lucy's weight function \cite{b12,b13}:
$$
w^{1D}(r<1) = (5/4h)(1-r)^3(1+3r) \ ; \ r \equiv |x|/h \ .
$$
$$
w^{2D}(r<1) = (5/\pi h^2)(1-r)^3(1+3r) \ ; \ r \equiv \sqrt{x^2+y^2}/h \ .
$$
The averages are not significantly different to those computed with
Hardy's more cumbersome approach\cite{b14}.
%Figure 2 goes here
\begin{figure}
\includegraphics[height=6cm,width=6cm,angle=-90]{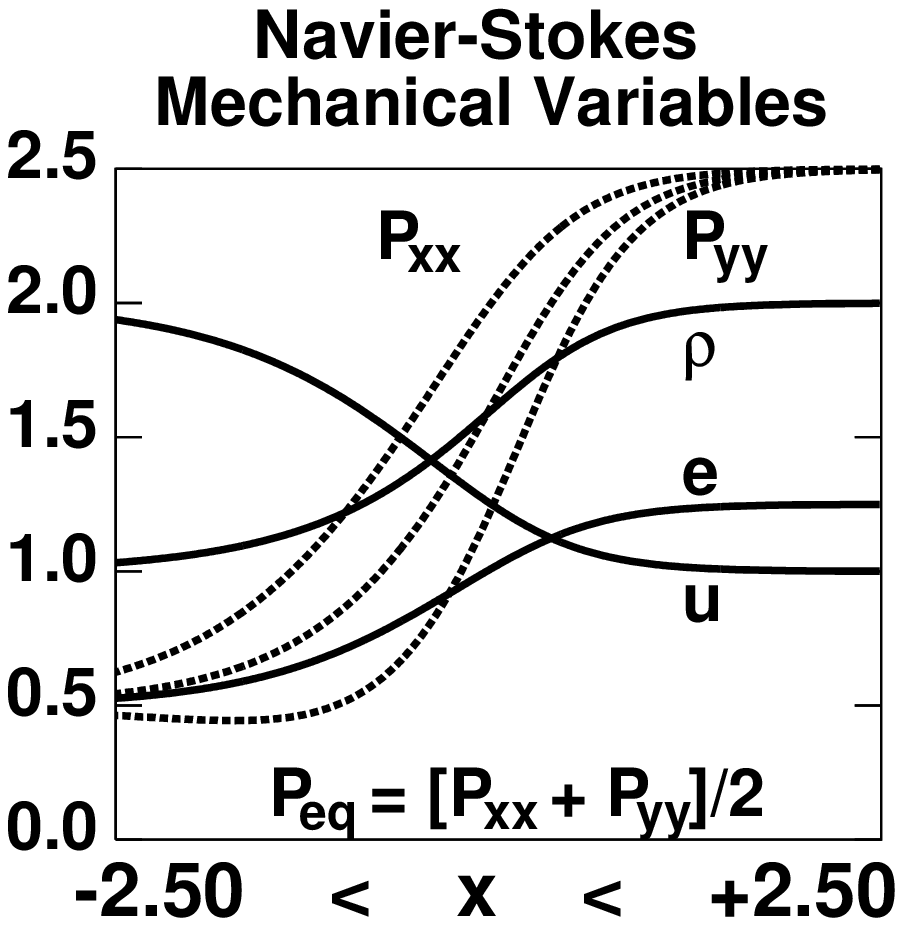}
\caption{
The nonzero pressure-tensor components from the Navier-Stokes equations
solution, $P_{xx}$ and $P_{yy}$, as well as
their average, $P = (\rho^2/2) + \rho T$, are shown as dashed lines along
with the velocity, energy, and density profiles.  In
this simple example it is assumed that the bulk viscosity vanishes so 
that the average of the longitudinal and transverse components is equal
to the equilibrium pressure.
}
\end{figure}

%Figure 3 goes here
\begin{figure}
\includegraphics[height=6cm,width=6cm,angle=-90]{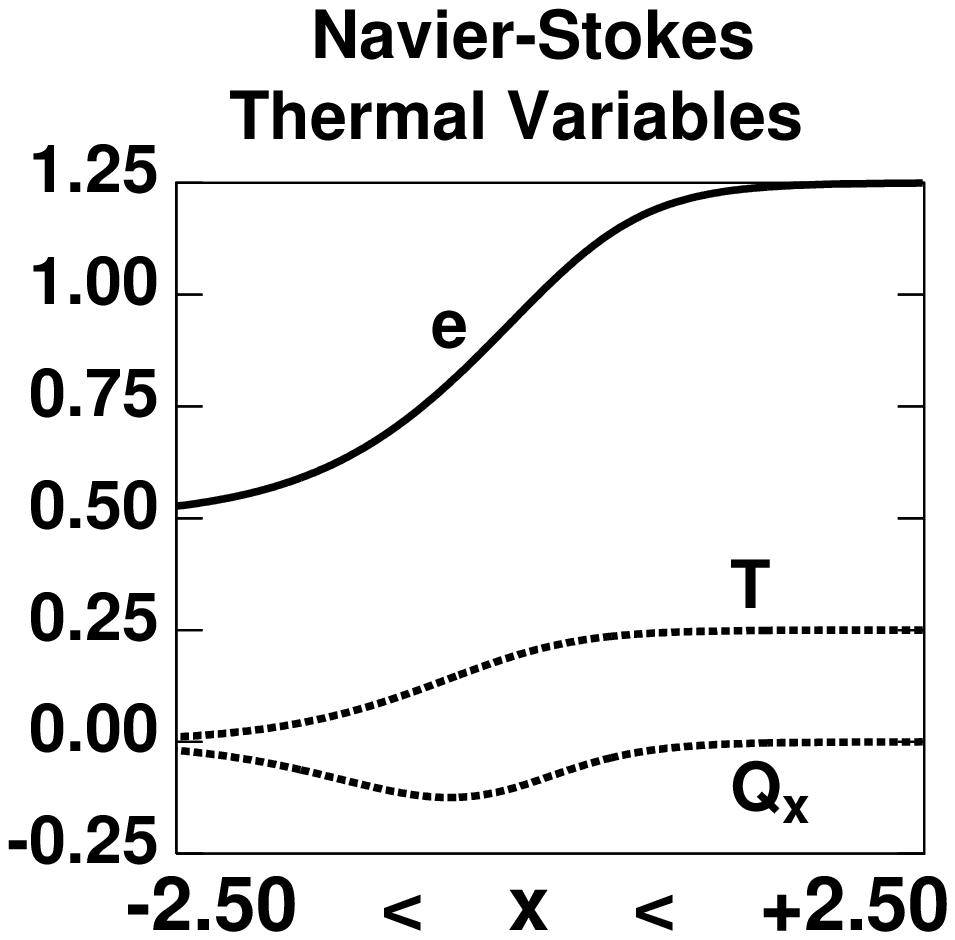}
\caption{
The energy, (scalar) temperature, and heat flux vector from the Navier-Stokes
equations are shown here.  The heat 
conductivity and shear viscosity coefficients were assumed equal to
unity in the underlying calculation.
}
\end{figure}

A preview of the one-dimensional averages results from molecular dynamics' simplest
continuum analog, a solution of the stationary Navier-Stokes equations.  For
simplicity, in the Navier-Stokes analog we use the constitutive relations
for the van-der-Waals-like model with shear viscosity and heat
conductivity of unity:
$$
P = \rho e = (\rho ^2/2) + \rho T \ ; \ e = (\rho /2) + T
$$
$$
(P_{xx} - P_{yy})/2 = - du/dx \ ; \ (P_{xx} + P_{yy})/2 = P \ ; \
$$
$$
Q_x = -dT/dx \ .
$$
This model is similar to our microscopic simulation model, but has a nonzero
initial pressure and energy.  A set of self-consistent cold and hot boundary
conditions for the Navier-Stokes velocity, pressure, energy, and scalar
temperature is as follows:
$$
u: [2 \rightarrow 1] \ ;
 \ \rho: [1 \rightarrow 2] \ ;
P: \ [1/2 \rightarrow 5/2] \ ;
e: [1/2 \rightarrow 5/4] \ ;
T: [0 \rightarrow 1/4] \ .
$$
These boundary conditions satisfy conservation of mass, momentum, and energy.
The (constant) mass, momentum, and energy fluxes {\em throughout} the shockwave
(not just at the boundaries) are:
$$
\rho u  = 2 \ ;
\ P_{xx}  + \rho u^2 = 5/2 \ ;
\ \rho u[e + (P_{xx}/\rho) + (u^2/2)] + Q_x = 6 \ .
$$
The most noteworthy feature of the numerical solution is the slight decrease
of $P_{yy}$ below the equilibrium value on the cold side of the shock.  Figure
2 shows the mechanical variables and Figure 3 the thermal variables near the
center of the shock as computed from the Navier-Stokes equations\cite{b11}.
A serious shortcoming of the Navier-Stokes equations is their failure to
distinguish the longitudinal and transverse temperatures.

\section{Averaged Results from Molecular Dynamics}

%Figure 4 goes here
\begin{figure}
\includegraphics[height=6cm,width=6cm,angle=-90]{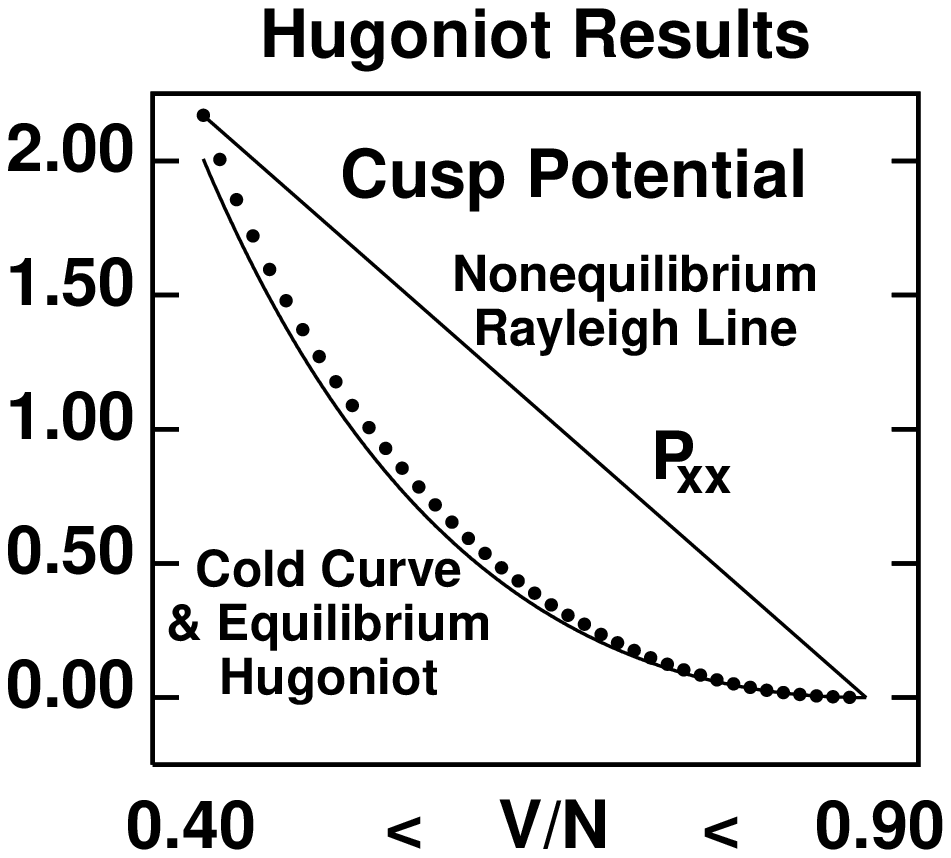}
\caption{
The longitudinal pressure tensor component $P_{xx}$ varies linearly
with volume, and follows the Rayleigh line.  The cold curve corresponds
to the pressure of a perfect static triangular lattice.  The equilibrium
Hugoniot, indicated by dots, corresponds to thermodynamic equilibrium
states accessible from the initial cold state by shockwave compression.
For $P_{yy}$ see Figure 7.
}
\end{figure}

%Figure 5 goes here
\begin{figure}
\includegraphics[height=8cm,width=8cm,angle=-90]{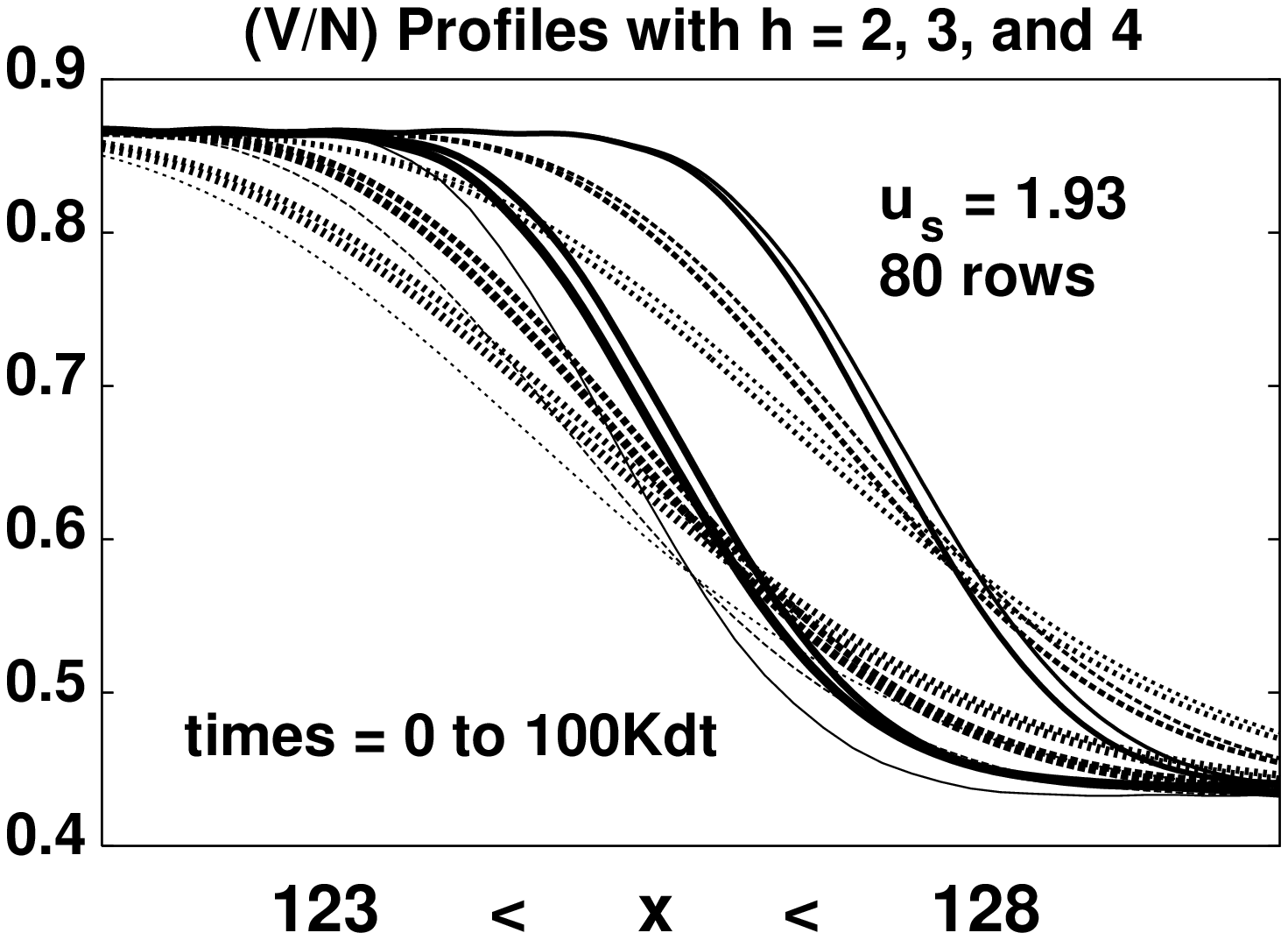}
\caption{
Shockwave profiles at five equally-spaced times.  The five line widths
correspond to times of 0, 25,000$dt$, 50,0000$dt$, 75,000$dt$, and 100,000$dt$.
The steepest profiles correspond to $h=2$.  Results for $h=3$ and $h=4$
are also shown.  The fourth-order Runge-Kutta timestep, here and throughout,
is given by $dt = 0.02/u_s \simeq 0.01$.
}
\end{figure}

One-dimensional averages reproduce the linear dependence of $P_{xx}$ on
the volume very well.  That linear dependence is the ``Rayleigh Line'',
shown in Figure 4.  The ``cold curve'' in that Figure is the calculated
pressure for a cold triangular lattice:
$$
P_{\rm cold}V = 3NrF(r) ; \ r = \sqrt{(V/V_0)} \ ; \ V_0 = \sqrt{3/4}N \ ; \
F(r) = (30/\pi)(1-r)^2 \ .
$$
That pressure lies a bit below the Hugoniot curve (the locus of all
equilibrium states which can be reached by shocking the initial state).
The Hugoniot pressure at each volume was generated by trial-and-error
isothermal (isokinetic) molecular dynamics runs, leading to the
temperatures satisfying the Hugoniot relation:
$$
E_{\rm hot} - E_{\rm cold} = +\Delta V[P_{\rm hot} + P_{\rm cold}]/2 \ ; \
\Delta V = V_{\rm cold} - V_{\rm hot} \ .
$$

Figure 5 shows typical one-dimensional snapshots of the shockwave profile,
$V(x)$.  The averages shown in the Figure were computed at 5 equally-spaced
times, separated by 25,000 timesteps.  The fluctuating motion of the
shockwave, of order unity in 100,000 timesteps, corresponds to fluctuations
in the averaged shock velocity of order 0.001.

The apparent shockwidth is sensitive to the range of the weighting function
$h$.  $h=2$ is evidently {\em too small}, as it leads to discernable wiggles in
the profile. The wider profiles found for $h=3$ and $h=4$ indicate that the
constitutive relation describing the shockwave must depend explicitly on $h$.
That is, $h$ must be chosen sufficiently large to avoid unreasonable
wiggles, but must also be sufficiently small to capture and localize the
changes occurring within the shockwave.

Two-dimensional averages are no more difficult to evaluate.  The density
at a two-dimensional gridpoint, for instance, can be evaluated by summing
the contributions of a few dozen nearby particles:
$$
\rho _r \equiv \sum _jw^{2D}_{rj} = \sum _jw^{2D}(|r-r_j|) \  .
$$
Such sums are automatically continuous functions of the gridpoint location
$r$. They necessarily have continuous first and second derivatives too,
provided that the weight function has two continuous derivatives, as
does Lucy's weight function\cite{b12,b13}.  Linear
interpolation in a sufficiently-fine grid can then provide {\em contours}
of macroscopic variables.  Figure 6 is an illustration, and shows the
contour of average density at 10 equally-spaced times.  The boundary
value of $u_s$ for that Figure was chosen as 1.92 rather than the
shock velocity of 1.93.  Thus the shockfront moves slowly to the left in
the Figure, with an apparent picture-frame velocity of $- 0.01$.

%Figure 6 goes here
\begin{figure}
\includegraphics[height=12cm,width=7cm,angle=-90]{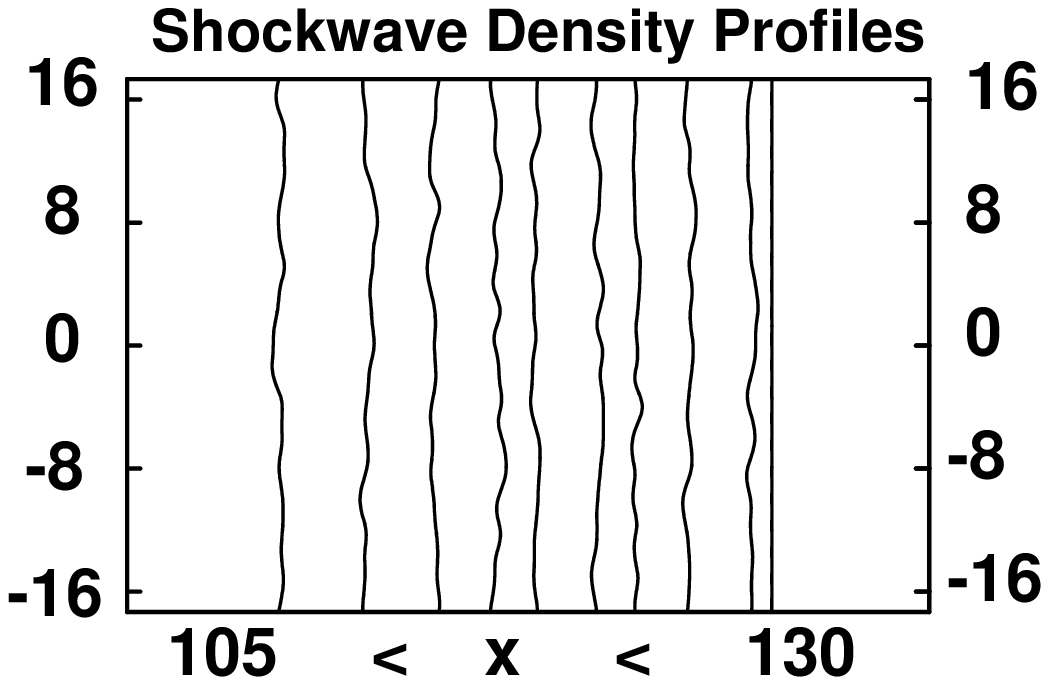}
\caption{
Ten average-density contours (corresponding to the shockfront position)
at equally-spaced times for a 40-row system.  The amplitude of the
fluctuations, of order unity,
is similar to the range of the weight function $h$.  The total timespan is
225,000  timesteps.  Because the entrance velocity, at $x=0$, is
$u_s$ = 1.92, rather than 1.93, the shockfront moves slowly toward the left,
with a picture-frame velocity of about -0.01. 
}
\end{figure}

Figure 7 shows the nonequilibrium
equation of state within the shockwave, the variation of the pressure
tensor components $P_{xx}$ and $P_{yy}$ with the specific volume,
$(V/N) \equiv (1/\rho)$.  $P_{xx}$ is insensitive to the smoothing length
$h$ (as is required by the momentum conservation condition defining the
Rayleigh line) while $P_{yy}$ shows a slight dependence on $h$.  This
lack of sensitivity of the pressure tensor suggests that nonequilibrium
formulations of the equation of state within the shock can be successful.

%Figure 7 goes here
\begin{figure}
\includegraphics[height=10cm,width=6cm,angle=-90]{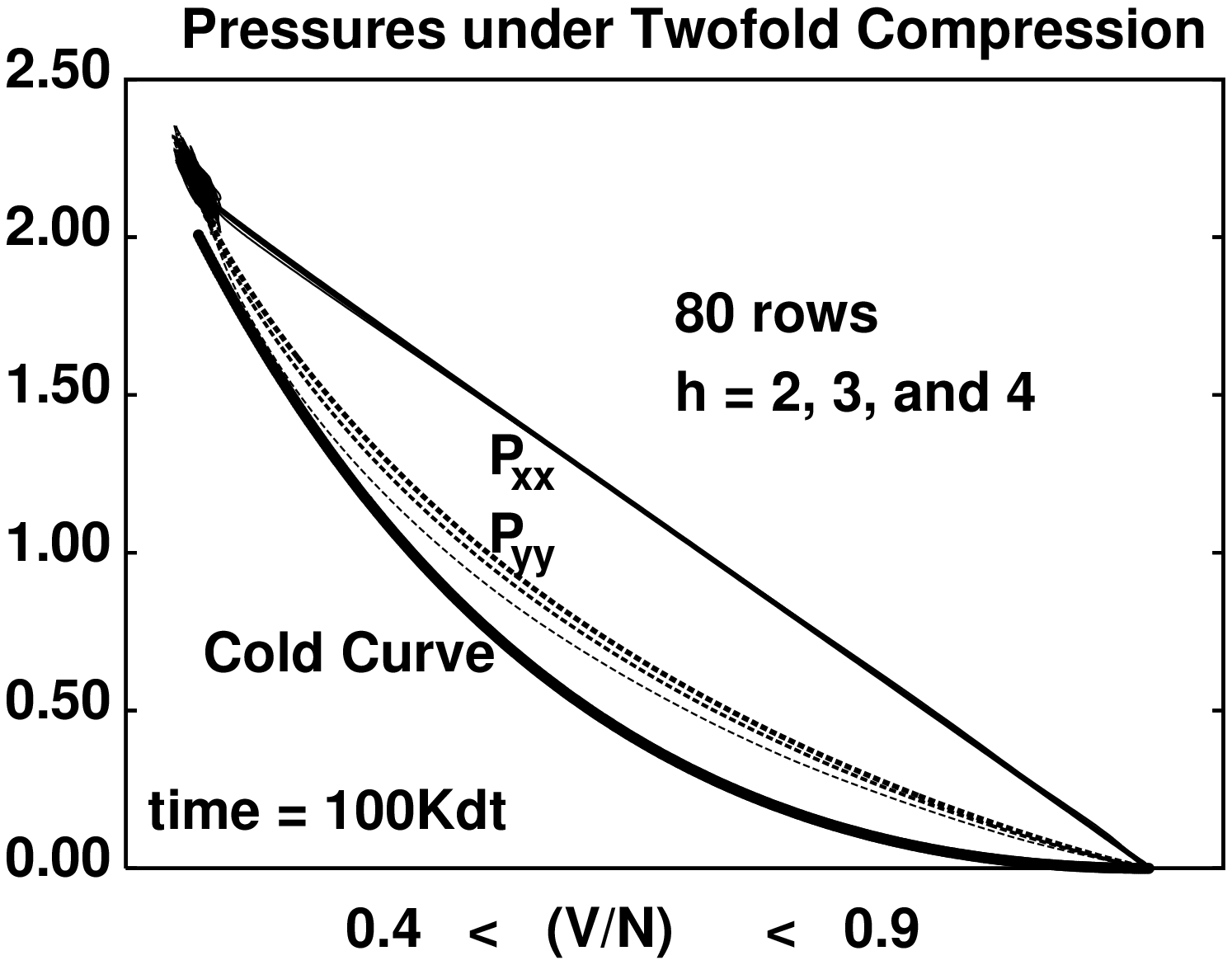}
\caption{
Typical snapshot of the dependence of $P_{xx}$ and $P_{yy}$ on the
volume $(V/N)$. The line width increases with the range $h = 2,$ $3,$ and $4$.
The range-dependence of $P_{xx}$ is too small to be seen here while it
is possible to see a small increase in $P_{yy}$ with increasing $h$.
}
\end{figure}

  The conventional
Newton and Fourier constitutive relations require that gradients be
examined too.
{\em Gradients} can be evaluated directly from sums including $\nabla w$.
Consider the gradient of the
velocity at a gridpoint $r$ as an example (where $w_{rj}$ is the weight
function for the distance separating the gridpoint from Particle $j$):
$$
(\rho \nabla \cdot v + v \cdot \nabla \rho)_r \equiv \nabla \cdot (\rho v)
\equiv \nabla_r \cdot \sum _jw_{rj}v_j = \sum_j v_j\cdot \nabla _rw_{rj} \ .
$$
Using the identity,
$$
\rho = \sum_j w_{rj} \ ,
$$
gives
$$
(\rho \nabla \cdot v)_r = \sum_j (v_j - v_r) \cdot \nabla_r w_{rj} \ .
$$
The tensor temperature gradient can be evaluated in the same way:
$$
(\rho \nabla \cdot T)_r = \sum_j (T_j - T_r)\cdot \nabla_r w_{rj} \ .
$$

Figure 8 compares the velocity gradients as calculated using three values
of $h$ to the pressure tensor using the same three values. We see that the
velocity
gradient is much more sensitive than is the stress to $h$, suggesting a
sensitive dependence of the Newtonian viscous constitutive relation on the
range of the weight function.  The data in the Figure indicate a shear
viscosity of the order of unity.  Gass' Enskog-theory viscosity\cite{b15}
confirms this estimate.

Figure 9 shows the temperature gradients.  There are two of these for each
$h$ because the longitudinal and transverse temperatures differ.  Again the
magnitudes of the gradients are relatively sensitive to $h$ while the
maximum in the nonequilibrium flux $Q_x$ is less so.  Again the heat
conductivity from the data is of the order of Gass' Enskog-theory estimate.

%Figure 8 goes here
\begin{figure}
\includegraphics[height=10cm,width=6cm,angle=-90]{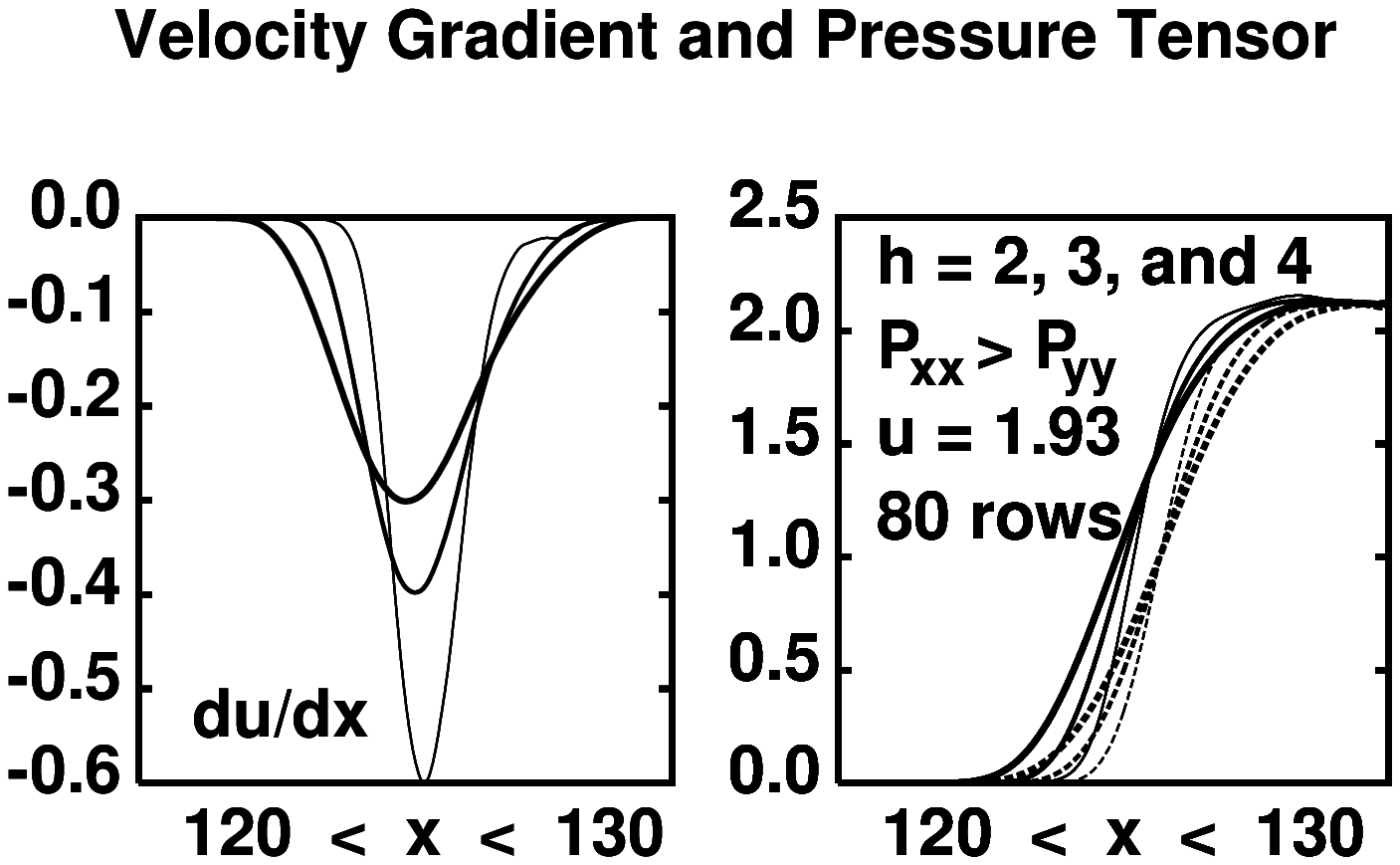}
\caption{
The velocity gradient, for three values of $h$, is much more sensitive than
is the shear stress, $(P_{yy}-P_{xx})/2$, to $h$.  The gradient extrema,
at 125.35, 125.18, and 125.01 precede the shear stress extrema at 125.67,
125.61, and 125.57 for $h =$ 2, 3, and 4.
}
\end{figure}

%Figure 9 goes here
\begin{figure}
\includegraphics[height=10cm,width=6cm,angle=-90]{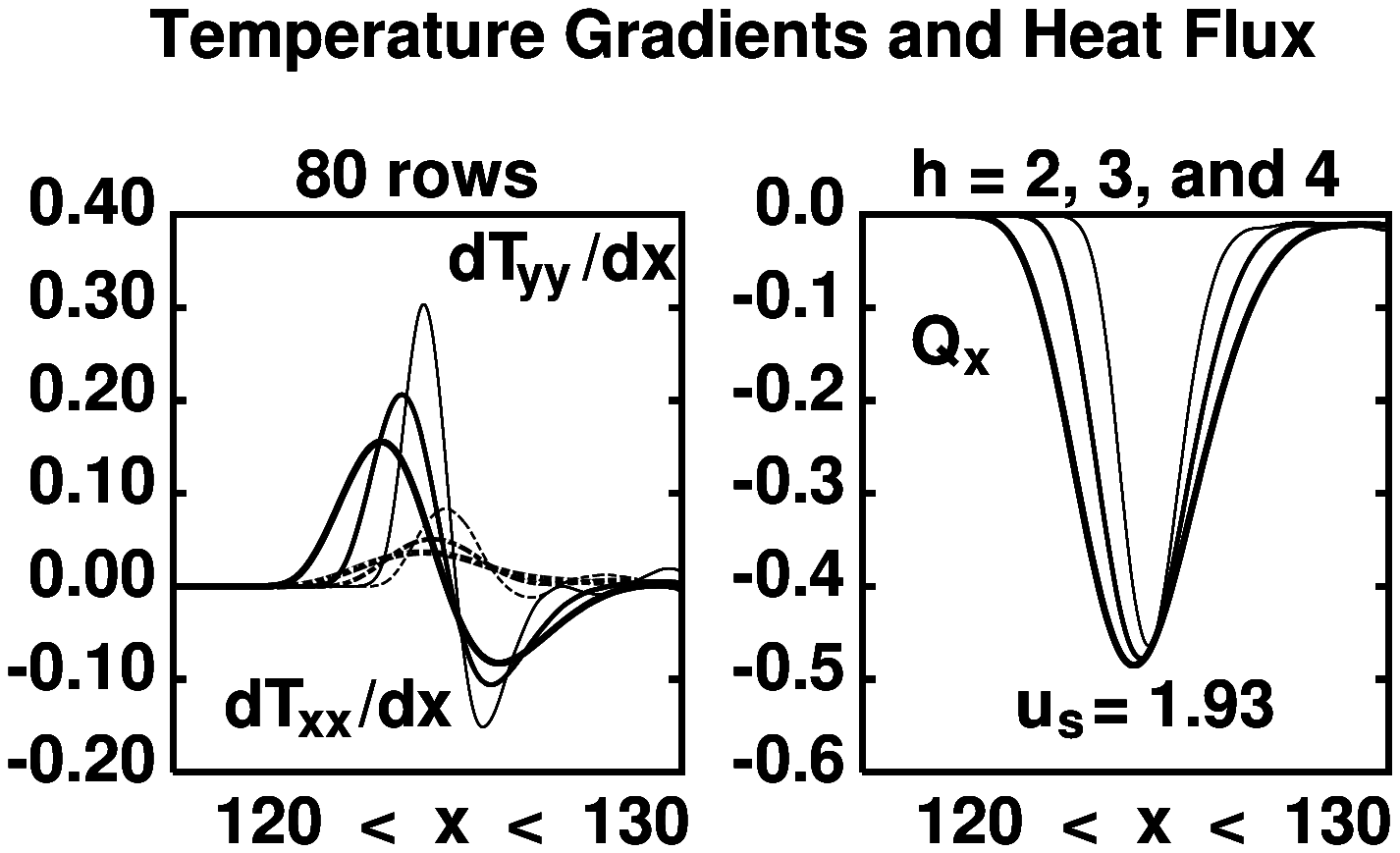}
\caption{
The temperature gradients (the dashed lines correspond to the transverse
temperature) for three values of $h$ correspond to the heat fluxes found
with the same $h$ values.  The pronounced maximum in $T_{xx}$ indicates a
violation of Fourier's Law, as the heat flux does not show a corresponding
change of sign.  The maxima in $dT_{xx}/dx$ occur at distances 124.93,
124.49, and 124.09, significantly leading the flux maxima at 125.44, 125.29,
and 125.15 for $h =$ 2, 3, and 4.
}
\end{figure}

\section{Conclusion: Failure of Navier-Stokes Equations}

Results from earlier shockwave simulations\cite{b9,b11,b16} have uniformly been
described as showing ``good'' or ``fairly good'' agreement with continuum
predictions.  Examining the more nearly accurate profiles made possible
with improved averaging techniques shows that the agreement is actually
limited, and in a qualitative way.
A more detailed look at the data shown in Figures 8 and 9 reveals a
consistent ``fly in the ointment'' pattern: the largest fluxes are
{\em not} located at the largest values of the gradients.  The fluxes
lag behind the gradients by a (relaxation) time of order unity.  This
shows that no simple instantaneous relationship links the fluxes to
the gradients. {\em In molecular dynamics the instantaneous stress cannot be
proportional to the instantaneous strain rate.}

The reason for this apparent contradiction of linear
transport theory is plain enough: the underlying molecular dynamics is
time-reversible, so that pressure is necessarily an {\em even} function
of velocity and time.  This same symmetry must be true also of any
spatially-averaged instantaneous pressure.
Because there is no possibility to find an instantaneous irreversible
constitutive relation with time-reversible molecular dynamics, it is
apparent that any attempt to ``explain'' local molecular dynamics
averages through irreversible macroscopic constitutive relations is
doomed to failure.

There is of course no real difficulty in carrying out the instantaneous
averages, in one or
two or three space dimensions, for today's molecular dynamics simulations.
On the other hand, the gap between the microscopic and macroscopic pictures
becomes an unbridgable chasm when the detailed spatiotemporal contradictions
between the two approaches are considered.

\section{Prospects}

The prospect of understanding shockwaves in gases has stimulated studies
of dilute gases, based on the Boltzmann equation\cite{b17,b18,b19,b20}.
Leo has been a driving force for this work.
Though the analysis is highly complex\cite{b19,b20} it has become
apparent that the Boltzmann equation is itself nicely consistent with
corresponding solutions using molecular dynamics\cite{b17,b18}, up to
a Mach number $M = u_s/c_{\rm cold}$ of 134. The
applicability of the Burnett equations, which include all second-order
contributions of the gradients to the fluxes, is still in doubt for
strong shockwaves in dilute gases\cite{b17,b18}.

Dense fluids will require a
new approach. Local averages must be defined.  Longitudinal and transverse
temperatures
must be treated separately.  The causal timelag between the forces
(velocity and temperature gradients) and the resulting momentum and
heat fluxes must be included in the modeling.  Although these
challenges are enormous, today's fast computers place the responsibility
for successfully meeting them squarely on physicists' imaginations.  The
excuse that the problem is too hard to tackle is no longer valid.  We can
look forward to many more contributions from Leo, his coworkers, and those
inspired and stimulated by his work.

\section{Acknowledgment}

We thank Paco Uribe, Michel Mareschal, Leopoldo Garc\'ia-Col\'in, and
Brad Holian for their comments on an earlier draft of this work.
Paco Uribe kindly exhumed and resurrected the Burnett solutions of
References 19 and 20 and established that there is indeed a timelag
between the Burnett forces and fluxes in dilute-gas shockwaves for
hard-sphere Mach numbers of both 2 and 134.  This seems to contradict
the lower part of Figure 2 in Reference 17.

\end{document}